\documentstyle[sprocl]{article}

\input{psfig}

\def\dslash{\not{\hbox{\kern-2pt $\partial$}}}
\def\eslash{\not{\hbox{\kern-2pt $\epsilon$}}}
\def\Dslash{\not{\hbox{\kern-4pt $D$}}}
\def\Aslash{\not{\hbox{\kern-4pt $A$}}}
\def\Qslash{\not{\hbox{\kern-4pt $Q$}}}
\def\Wslash{\not{\hbox{\kern-4pt $W$}}}
\def\pslash{\not{\hbox{\kern-2.3pt $p$}}}
\def\kslash{\not{\hbox{\kern-2.3pt $k$}}}
\def\qslash{\not{\hbox{\kern-2.3pt $q$}}}

\def\NPB#1{{\em Nucl.~Phys.~\bf B#1}}
\def\PLB#1{{\em Phys.~Lett.~\bf B#1}}

\def\PRD#1{{\em Phys.~Rev.~\bf D#1}}
\def\ZPC#1{{\em Z. Phys.~\bf C#1}}

\def\be{\begin{equation}}
\def\ee{\end{equation}}
\def\bea{\begin{eqnarray}}
\def\eea{\end{eqnarray}}

\begin{document}

\title{$o(\alpha_s)$ EFFECTS IN t\=t PRODUCTION AT THE NLC ABOVE THRESHOLD
\footnote{Presented at PASCOS 98, Boston, MA, March 1998;
to appear in the Proceedings of PASCOS 98 (World Scientific).}}

\author{GEORGE SIOPSIS}

\address{Department of Physics and Astronomy \\ The University of Tennessee \\
Knoxville, TN 37996--1200 \\ USA\\E-mail: gsiopsis@utk.edu}


\maketitle\abstracts{
We study the effects of gluon interference in the production and semi-leptonic
decay of a t\=t pair above threshold at the Next Linear Collider (NLC).
We calculate all matrix elements to next-to-leading order and use the resulting
expressions for the development of a Monte Carlo event generator.
Our results show effects at the level of 10\% in differential cross-sections.
We thus extend previous results obtained by analytical
means in the soft-gluon limit.
}


Due to its
large mass,
and, consequently, its large decay width ($m_t = 176$ GeV,
$\Gamma_t = 1.57$ GeV),
the top quark
decays before it has a chance to
hadronize. 
It is therefore possible to perform a precision analysis
of top quark production and decay, which facilitates the identification of effects attributable
to new physics beyond the Standard Model.
The best laboratory for such explorations is an $e^+e^-$ collider, because the initial state
is free of color effects.
Such a machine (the Next Linear Collider (NLC)) will hopefully be
built in the not-so-distant future. 
Detailed studies of processes through which one may extract
useful information have already been performed~\cite{ref5a}. 
Gluon radiation contributes at the 10\% level,
in general. Although it does not modify inclusive cross-sections, it has a non-negligible effect
on various final-state distributions. Therefore, it has to be included in a precision study of
top quark physics. In studying gluon radiation, one cannot ignore the effects of gluon interference.
The latter depends upon the value of $\Gamma_t$ and may be an important tool in the study of
this parameter, in addition to precision studies~\cite{ref6}.

We have studied gluon interference effects in the process
\begin{equation}
  \label{eq0}
  e^+e^- \to \gamma^\star \;,\; Z^\star \to t\bar t \to W^+b\; W^-\bar b \; g\to
\ell^+\nu b \; \ell^-\bar\nu\bar b\; g
\end{equation}
We have calculated all amplitudes
analytically, with the aid of the symbolic manipulation package FORM~\cite{ref7}. The resulting expressions
were then used for the development of a Monte Carlo event generator. The generator was based on
the C++ program written by Schmidt for the study of the same process~\cite{ref8}, in the
narrow top width approximation. Details of the calculation are reported elsewhere~\cite{ego}.
The results from our simulations are in
agreement with analytical results obtained by Khoze, {\em et al.}~\cite{ref9} in the soft gluon regime. They are also similar
to the results obtained by Peter and Sumino~\cite{ref10} in the threshold region, where color Coulomb effects dominate~\cite{ref11}.

At tree level, the amplitude for t\=t production and decay factorizes.
With polarized $e^+e^-$ beams, one can perform a clean helicity analysis,
because the spins of the top and anti-top quarks get transferred to the final
products. It is then advantageous to express the amplitudes in terms of helicity
states~\cite{ref8}.

Loop diagrams at $o(\alpha_s)$ include a pentagon diagram (gluon exchange
between the b and \=b quarks), which renders the calculation of
$o(\alpha_s)$ effects cumbersome. To systematically calculate these
effects, observe that all loop diagrams contain one or more internal t or \=t
legs, which are rapidly decaying quarks. To take advantage of this fact, we
define the quantities
\begin{equation}
  \label{eq7}
\eta_1 = (p_{\bar t}^2 - m_t^2)/m_t^2 \;,\quad
\eta_2 = (p_t^2 - m_t^2)/m_t^2 \;.
\end{equation}
The momenta of t and \=t are defined in terms of the quarks' final decay products.
$\eta_1$ and $\eta_2$ are small parameters,
$o(\Gamma_t/m_t)$. Physically, this means that t and \=t are produced
nearly on shell.
It is advantageous to expand the
amplitude in these parameters,
\begin{equation}
  \label{eq10}
{\cal A} (\eta_1,\eta_2) = {\cal A}\Big|_{\eta_1=\eta_2=0}
+\eta_1 \left. {\partial {\cal A}\over \partial\eta_1}
\right|_{\eta_2=0} + \eta_2 \left. {\partial {\cal A}\over
\partial \eta_2} \right|_{\eta_1=0} + o(\eta_{1,2}^2)\;.
\end{equation}
where ${\cal A}$ needs to include both virtual and real gluon
contributions (soft and collinear), in order to avoid singularities.


After calculating amplitudes analytically at first order in
$\eta_{1,2}$,
we used the formulas to construct a Monte Carlo event generator.
The generator is based on a program written by Schmidt in C++. Schmidt's program generates events at tree
level without gluon production with 100 \% efficiency. This is achieved by making use of helicity states, not only for the final
states, but also for the intermediate t and \=t quark states. The amplitude is then expressed in terms
of helicity angles which are defined in different Lorentz frames for the three vertices, respectively.
This construction is carried over to one-loop level, if one neglects gluon interference effects, because
the amplitude is still factorizable at that level. The virtual, soft and collinear corrections amount
to corrections in the form factors for the three vertices in the diagrams.
Thus, one can construct a very efficient event generator.

To account for gluon interference effects, one may not rely on helicity angles, because the amplitudes
are no longer factorizable, and need to be expressed in terms of a single Lorentz frame. This complicates their
calculation. To simplify the expressions, we have introduced polarization vectors for the intermediate
$W^\pm$ boson states defined with respect to the b (\=b) quarks (which are treated as massless).
This quadruples the number of diagrams, but simplifies the final expressions.
At tree level, we have checked the accuracy of our expressions numerically by using Schmidt's program.
At loop level, we have performed a similar check in the limit $\eta_{1,2} \to 0$.

Since gluon interference effects are relatively small, we took advantage of Schmidt's efficient
code, by keeping the event generating procedure and introducing the new effects as a correction to the
weight.

We set the mass of the b quark to zero in matrix elements. This is essential for
the simplification of the expressions for the various amplitudes. We also neglected effects of
$o(\eta_{1,2}^2)$, which include gluon interference between the b and \=b quarks. Such effects are
very small and are not likely to contribute at the desired accuracy for the NLC. Should the need for higher
accuracy arise, our method can be readily applied to accommodate these effects. We also did not couple
the program with a jet algorithm. This would be necessary for a direct comparison with experimental data.
Initial-state radiation effects~\cite{ref12} were included in the program, but will not be shown here.

\begin{figure}[t]
\rule{0.5cm}{0mm}
\psfig{figure=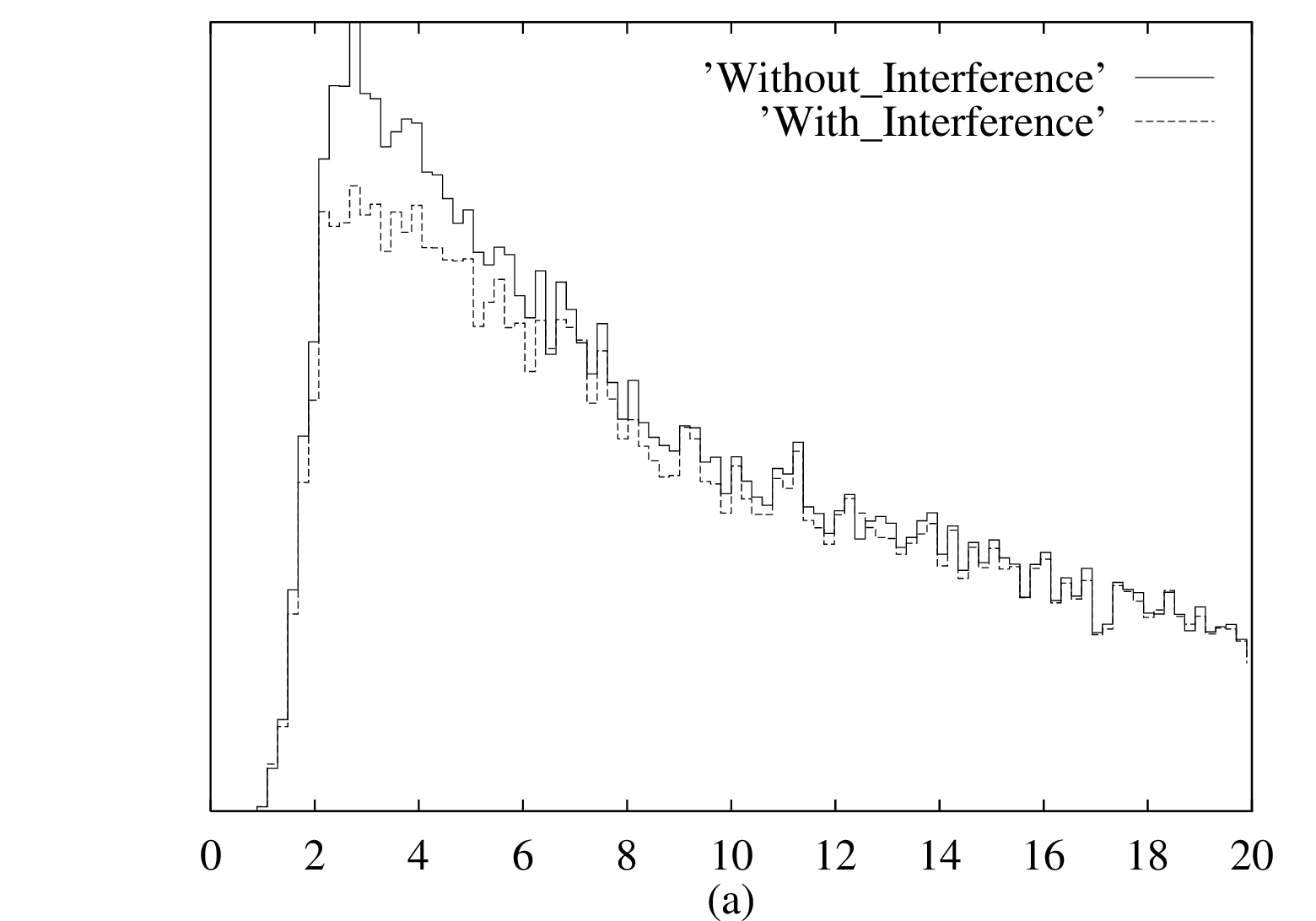,width=5cm,height=1.5in}\rule{1cm}{0mm}\psfig{figure=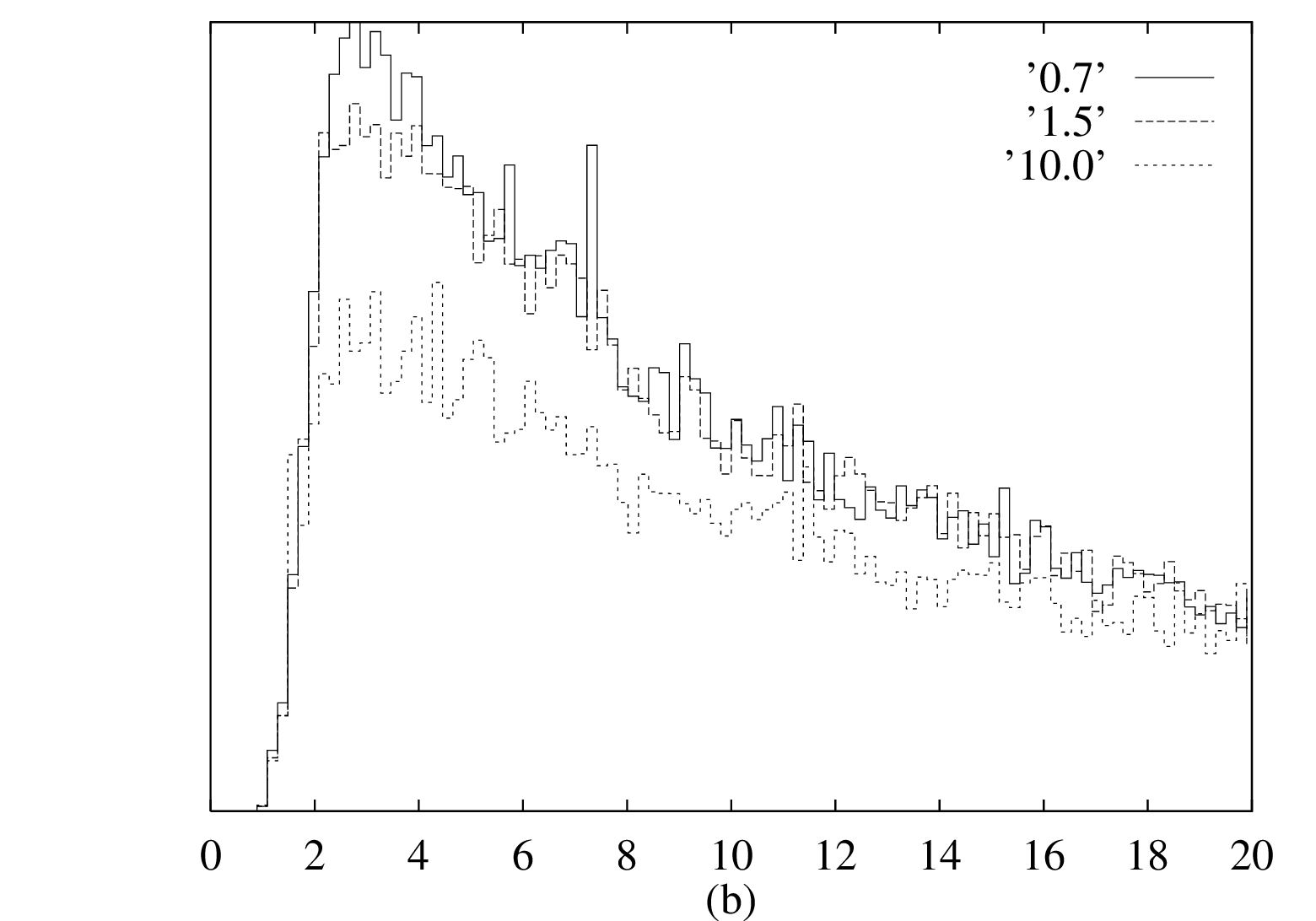,width=5cm,height=1.5in}
\caption{Distribution of gluon energy at center-of-mass energy of
400~GeV. In {\em (a)}, we exhibit interference effects. In {\em (b)}, we show the
dependence on $\Gamma_t$. \label{fig4}}
\end{figure}
Next, we present numerical results obtained with the Monte Carlo event generator. As has been shown, there
is a complete cancellation between virtual and real gluon contributions to an {\em inclusive}
cross-section~\cite{ref9a}. Moreover, differential cross-sections receive contributions from
gluon interference effects to $o(\alpha_s \Gamma_t/m_t)$. The major contribution to
such corrections comes from soft gluons of energy $E_g \sim \Gamma_t$~\cite{ref9a}.
Our results include hard gluon effects and are in general agreement with the
semi-classical analysis of soft gluon production~\cite{ref9,ref9a}.
Indicatively, we plot the distribution of the gluon energy (Fig.~\ref{fig4})
at center-of-mass energy $\sqrt s = 400$~GeV.
We have introduced a jet resolution parameter $\eta_{cut} = 0.05$
by demanding
$(k+p_b)^2/s ,(k+p_{\bar b})^2/s  > \eta_{cut}^2$
where $k^\mu$ is the four-momentum of the gluon.
We have imposed no other cut on the gluon.

In Fig.~\ref{fig4}{\em (a)}, we have compared distributions
with and without interference. 
In general, interference effects enter at a 10 \% level, so they need to be
included in a precision study of top quark production and decay.
Moreover, in Fig.~\ref{fig4}{\em (b)}, we show the
dependence of the distribution on the decay width of the top quark,
$\Gamma_t$. The decay width has a most
pronounced effect in the range of gluon energies
$E_g \sim \Gamma_t$~\cite{ref6}.
This shows that gluon interference effects in the soft gluon regime
can be used to measure the top quark decay width. This can already be seen
by a semi-classical analysis~\cite{ref6}, but the full quantum mechanical
calculation, which is done here, is needed for an accurate theoretical prediction.
Further discussion can be found elsewhere~\cite{ego}.

\section*{Acknowledgments}

{\small
I am indebted to Carl Schmidt for letting me use his program. I also wish to thank Linda Arvin and Eric
Conner for valuable help with the development of the program.
Research supported by the DoE under grant DE--FG05--91ER40627.
}

\end{document}